\documentclass{aa}  

\usepackage{graphicx}
\usepackage{txfonts, natbib}
\usepackage[]{hyperref}
\begin{document}

\title{The IAG solar flux atlas: Accurate wavelengths and absolute convective
  blueshift in standard solar spectra}


\author{A. Reiners\inst{1}
  \and
  N. Mrotzek\inst{1}
  \and 
  U. Lemke\inst{1}
  \and
  J. Hinrichs\inst{1}
  \and
  K. Reinsch\inst{1}
}

\institute{Georg-August Universit\"at G\"ottingen, Institut f\"ur Astrophysik, 
  Friedrich-Hund-Platz 1, 37077 G\"ottingen, Germany\\
  \email{Ansgar.Reiners@phys.uni-goettingen.de}
}

\date{Accepted Nov 9, 2015}

 
\abstract{We present a new solar flux atlas with the aim to understand
  wavelength precision and accuracy in solar benchmark data. The atlas
  covers the wavelength range 405--2300\,nm and was observed at the
  Institut f\"ur Astrophysik, G\"ottingen (IAG) with a Fourier
  Transform Spectrograph. In contrast to other FTS atlases, the entire
  visible wavelength range was observed simultaneously using only one
  spectrograph setting. We compare the wavelength solution of the new
  atlas to the Kitt Peak solar flux atlases and to the HARPS
  frequency-comb calibrated solar atlas. Comparison reveals
  systematics in the two Kitt Peak FTS atlases resulting from their
  wavelength scale construction, and shows consistency between the IAG
  and the HARPS atlas. We conclude that the IAG atlas is precise and
  accurate on the order of $\pm 10$\,m\,s$^{-1}$ in the wavelength
  range 405--1065\,nm while the Kitt Peak atlases show deviations as
  large as several ten to 100\,m\,s$^{-1}$. We determine absolute
  convective blueshift across the spectrum from the IAG atlas and
  report slight differences relative to results from the Kitt Peak
  atlas that we attribute to the differences between wavelength
  scales. We conclude that benchmark solar data with accurate
  wavelength solution are crucial to better understand the effect of
  convection on stellar RV measurements, which is one of the main
  limitations of Doppler spectroscopy at m\,s\,$^{-1}$ precision.  }


   \maketitle
%

\section{Introduction}

High resolution spectra of the Sun are important benchmarks for
various astrophysical fields. The high quality in spectral resolution
and signal-to-noise ratio (SNR) allows investigation of fundamental
processes in atomic and molecular physics, the structure of the solar
atmosphere, and the subtleties of line formation including convective
motion \citep[see, e.g.,][]{1982ARA&A..20...61D}. The latter is
particularly important for our understanding of stellar spectra
\citep[e.g.,][]{2009ApJ...697.1032G}, for which observations of the
Sun as a star -- solar flux observations -- are required. Another
important field for solar benchmark data is the growing interest in
ultra-high precision radial velocity observations (m\,s$^{-1}$ and
even cm\,s$^{-1}$ precision) for which stellar convection becomes
relevant as a fundamental limit \citep{1997ApJ...485..319S,
  2013A&A...551A.101M, 2014ApJ...796..132D}. Reaching out to
m\,s$^{-1}$ Doppler precision and below requires that the wavelength
scale of solar benchmark data is precise and accurate at that
level. While precision is enough when data taken with the same
instrument and the same star are analyzed, e.g., for radial velocities
(RV) of one star, a comparison between data from other stars to solar
data, taken with different instruments and from different parts of the
Sun (full disk, disk center, spot, etc.) requires an understanding of
the accuracy of the wavelength scales. In this paper, we present a new
high fidelity atlas of the solar flux and analyze precision and
accuracy of standard solar atlases.

A comprehensive overview about the lines asymmetries and wavelength
shifts in solar and stellar data was provided by
\citet{2008A&A...492..199D}. His Fig.\,5 shows the remarkable
difference of a few 100\,m\,s$^{-1}$ between Ti\,\textsc{ii} bisectors
when calculated from different solar disk center spectra. We focus on
solar flux atlases, i.e., spectra taken from the entire solar disk or
the Sun observed as a star that can be used to compare stellar to
solar observations. The most commonly used solar flux atlas is the
Kitt Peak atlas by \cite{1984sfat.book.....K} observed at McMath
National Solar Observatory with a Fourier Transform Spectrograph
(FTS). Another flux atlas from this facility was published in
\cite{2011ApJS..195....6W}. It is often assumed that the FTS atlases
provide high wavelength accuracy but this is not necessarily the case
at a level of a few ten m\,s$^{-1}$. In this paper, we carry out a
detailed comparison between the two Kitt Peak FTS atlases and the new
atlas presented here. As third benchmark, we include the solar flux
atlas from \cite{2013A&A...560A..61M} observed with HARPS and
wavelength calibrated with a Laser Frequency Comb (LFC). While the
spectral fidelity of this atlas is not as high as in the FTS atlases
(because of lower instrumental resolution) and the wavelength coverage
is limited, the wavelength solution is presumably the most accurate
currently available.

\begin{table*}
  \caption{\label{tab:settings}FTS settings used for the two solar atlases. 
    Configurations are defined by the beam-splitter (BS), the detector, and the maximum
    resolving power set by the arm length used for the scan. Column 6
    shows values of the resolving power in terms of $\lambda /
    \Delta\lambda$. The resolution is limited to $R < 10^6$ by optical
    elements and alignment.}
  \centering              
  \begin{tabular}{c r c c r c}
    \hline\hline          
    \noalign{\smallskip}
    & Range [nm] & BS & Detector & Res. power & $\lambda / \Delta\lambda$ [$10^6$] \\
    \noalign{\smallskip}
    \hline
    \noalign{\smallskip}
    VIS & 405 -- 1065 & Quartz & Si & 0.01\,cm$^{-1}$ & 2.47 -- 0.94  \\
    NIR & 1000 -- 2300 & CaF$_2$ & InSb &  0.0044\,cm$^{-1}$ & 2.27 -- 0.99  \\  
    \noalign{\smallskip}
    \hline\hline
  \end{tabular}
\end{table*}

The accuracy that can be reached in the wavelength calibration of
solar atlases is a crucial factor for the future development of
high-precision RV surveys. At the 1\,m\,s$^{-1}$ level, we observe the
influence of stellar convection and activity, and stellar RV surveys
can be fooled or at least hampered by the presence of regular changes
in stellar convective patterns. One of the most challenging tasks
today for our understanding (and correction) of Doppler shifts from
convective patterns is the effect of suppressed convective blueshift
in active regions
\citep{1981A&A....96..345D}. \citet{2014ApJ...796..132D} provided a
tool for simulating the influence of activity on stellar RV
measurements. Useful templates for this task are FTS measurements from
the solar disk and solar spots. However, the results from such a
simulation need to be interpreted with care because the absolute
wavelength scales of the template spectra are not understood to the
level required for precision RV work. It is the purpose of this paper
to investigate the typical wavelength uncertainties of the available
standard solar atlases, and to identify the challenges we have to
overcome in order to provide useful input for future RV simulations.

The paper consists of mainly three parts: first, in
Sects.\,\ref{sect:Instrument}, \ref{sect:Observations}, and
\ref{sect:DataReduction} we describe our new FTS atlas and the other
atlases used for comparison. Second, we compare the wavelength
solutions of the atlases among each other in
Sect.\,\ref{sect:Comparison}. Finally, we use our new atlas in
Sect.\,\ref{sect:Blueshift} to re-investigate convective blueshift in
the solar flux spectrum, and we provide absolute values for convective
blueshift as a function of line depth.

\section{The Instrument}
\label{sect:Instrument}

At the Institut f\"ur Astrophysik, G\"ottingen (IAG), we are operating
a Fourier Transform Spectrograph (FTS) Bruker IFS 125 HR with a
maximum optical path length of 228\,cm in asymmetric mode (50\,cm in
symmetric mode), resulting in a maximum resolving power exceeding $R =
10^6$ for wavelengths shorter than 2\,$\mu$m, but the resolution is
practically limited to $R \approx 10^6$ by design of the optical
elements. The instrument is operated under vacuum at a pressure of
approx. 0.1\,hPa. To cover wavelengths from optical to infrared, we
utilize two different beam splitter and detector configurations. We
use a Quartz beam-splitter together with a Si detector for wavelengths
in the range 405--1065\,nm, and a CaF$_2$ beam-splitter with a InSb
detector for the range 1000--2300\,nm. The two modes cannot be
operated simultaneously so that two different spectra are taken that
overlap in the region 1000--1065\,nm.

Our FTS has two feeds, one standard entrance accepting a focused beam
and one entrance for a parallel beam. We are using the parallel beam
entrance to inject light from a fiber through a reflective collimator
feeding parallel light into the FTS. At the other side of the fiber,
light from the Sun is coupled in also using a 12.5\,mm reflective
collimator. The collimator is located after the second mirror of a
siderostat\footnote{http://www.uni-goettingen.de/en/217813.html}
installed at the roof of our institute building. The siderostat
mirrors are flat and used for tracking only. The collimator creates an
image of the solar disk at the fiber entrance, the image diameter is
approximately 2/3 of the fiber diameter leaving room for tracking
inaccuracies. Tracking can cause imperfect coupling that can
potentially lead to a loss of light from limb areas of the solar
disk. We found that this effect can have an amplitude of several
10\,m\,s$^{-1}$ between different observations, and we will provide a
detailed analysis of the effect in a forthcoming paper. For the solar
atlas, this can lead to imperfect centering of the individual spectra
before co-adding resulting in a slightly disturbed line profile and an
offset of the solar lines on an absolute scale. It would, however, not
influence the relative positions of spectral lines. A 20\,m long
800\,$\mu$m fiber is used to feed the light from the telescope into
the FTS.

\section{Observations}
\label{sect:Observations}

We obtained spectra with two different settings, one visual (VIS) and
one near-infrared (NIR). Details of the settings are given in
Table\,\ref{tab:settings}. Observations were carried out between March
and July 2014. A log of our observations is provided in
Table\,\ref{tab:observations}.  Switching between settings involves a
change of the beam-splitter for which the FTS must be opened leading to
loss of vacuum. Therefore, we did not observe the Sun in different
settings on the same day. Since the optical design of our FTS limits
the spectral resolution to $R \approx 10^6$, we chose to limit the
resolving power (set by the scan length of the FTS) to 0.01\,cm$^{-1}$
in the VIS setting while the highest resolving power,
0.0044\,cm$^{-1}$, was chosen for the NIR setting.

The noise of a spectrum observed with an FTS grows with the total
number of photons reaching the detector. This leads to the unfortunate
effect that while the signal for a spectrum in a given wavelength
range can only be provided by photons at that range, noise is
collected from all photons across the entire range the detector is
sensitive to. A way to reduce the noise is to limit the wavelength
range with optical filters. Large wavelength coverage can then be
achieved by stitching together spectra observed with different
filters. One disadvantage of this strategy is that the different
atlases have systematically different wavelength solutions (see next
Section). With the aim of providing an atlas with a consistent
wavelength solution across the largest possible wavelength range, we
chose to not use filters for noise reduction. The consequence is that
our spectra are relatively noisy and we require a large number of
scans to produce a high signal-to-noise atlas.

We obtained spectra in packages of 10 scans each that were averaged by
the instrument software. Obtaining a package of 10 scans took approx.\
15\,min for the VIS and 17\,min for the NIR setting. The NIR scan took
longer because the scan length must be longer to achieve the higher
resolving power. In total, we obtained 1190 scans in the VIS and 1130
scans in the NIR setting. The dates at which spectra were taken are
shown in Table\,\ref{tab:observations}. Observations were carried out
over several months during which the solar activity changed
significantly. The final spectrum is therefore an average over
different levels of activity; the sunspot number as reported by
\cite{sidc} is provided in Table\,\ref{tab:observations} for each
observing day. Solar activity is known to distort line profiles by an
amount equivalent to a RV shift of a few m\,s$^{-1}$ similar to the
uncertainty in individual scans. Averaging spectra at different levels
of activity therefore has no significant influence on the final solar
atlas.

\begin{table}
  \caption{Observing log. Scans were taken in blocks of ten. Wavelength
    correction is given in units of $(1 + \kappa)$ with $\kappa \approx -\varv/c $.}
  \label{tab:observations}
  \centering              
  \begin{tabular}{c c c c c}
    \hline\hline          
    \noalign{\smallskip}
     & Date & \# of Scans & Sunspot Nr & $(1 + \kappa)$ \\
    \noalign{\smallskip}
    \hline
    \noalign{\smallskip}
VIS &2014 03 07 & 220 & 134 & 0.99999957 \\
    &2014 03 10 & 170 & 113 & 0.99999973 \\
    &2014 04 16 &  50 & 187 & 0.99999976 \\
    &2014 04 17 & 280 & 199 & 0.99999963 \\
    &2014 04 20 & 190 & 173 & 0.99999970 \\
    &2014 07 11 & 100 & 144 & 1.00000733 \\
    &2014 07 17 &  60 &   0 & 1.00000713 \\
    &2014 07 22 &  50 &  39 & 1.00000702 \\
    &2014 07 23 &  70 &  65 & 1.00000696 \\
    \noalign{\smallskip}
    &total & 1190\\
    \noalign{\smallskip}
    \hline 
    \noalign{\smallskip}
NIR &2014 02 24 & 150 & 130 & 0.99999979 \\  
    &2014 03 14 & 150 & 109 & 0.99999963 \\
    &2014 03 20 & 190 & 139 & 0.99999968 \\
    &2014 03 27 & 210 & 115 & 0.99999971 \\
    &2014 06 06 & 300 & 119 & 0.99999973 \\
    &2014 06 07 & 130 & 133 & 0.99999975 \\
    \noalign{\smallskip}
    &total & 1130\\
    \noalign{\smallskip}
    \hline\hline
  \end{tabular}
\end{table}

\section{Data Reduction}
\label{sect:DataReduction}

Data reduction of FTS data is very different from grating
spectrographs. We refer to the standard literature for an introduction
to Fourier Transform Spectroscopy \citep[e.g.,][]{2001ftsp.book.....D,
  2007ftis.book.....G}. A great advantage of Fourier Transform
Spectrographs is that they are free of instrumental scattered
light. This allows to accurately track the detailed shape of solar
spectral lines particularly at very high resolution. Furthermore, the
sharp line emission from a laser is used to electronically count the
wavenumbers while scanning; this technique provides a first-order
wavenumber scale with every scan. The wavenumber scale provided by the
laser is generally applicable to all wavelengths, which is another
great advantage of the FTS because all wavelength ranges truly see the
same optical path, although for technical reasons the laser itself
follows a slightly different path through the instrument. This
difference is linear in wavenumber
\citep{2004JOSAB..21.1543S}\footnote{We note that Salit el al.\ showed
  linearity to an accuracy of $6 \cdot 10^{-9} = 2$\,m\,s$^{-1}$ in
  their FTS in the wavelength range 320--670\,nm. We believe that the
  linearity of our wavelength scale is not significantly different.}
and can be accounted for by correcting the wavenumber scale \citep[see
Chapter 2.6 in][]{2007ftis.book.....G}
\begin{equation}
  \nu_{c} = \nu \ (1 + \kappa)
\end{equation}
with $\nu$ the uncalibrated wavenumber, $\nu_{c}$ the calibrated
wavenumber, and $\kappa$ the correction factor
($\kappa\approx-\varv/c$; with $c$ the speed of light and $\varv$ an
effective Doppler velocity). To determine $\kappa$, we measure the
positions of O$_2$ absorption lines imprinted by Earth's
atmosphere. These telluric lines are much narrower than the solar
lines and are not subject to Doppler shift from Earth's rotation or
orbit. They are stable on the level $\la$ 5--10\,m\,s$^{-1}$
\citep{1982A&A...114..357B, 2010A&A...511A..55F}. Correction factors
$(1 + \kappa)$ are determined for each day after averaging all scans
taken during that day, they are reported in
Table\,\ref{tab:observations}.

After determining the correction factors from daily averaged spectra,
we went back to the individual packages containing 10 scans each. To
each individual package, we first applied the $(1 + \kappa)$
correction. After that, each package was corrected for the Doppler
shift caused by the relative motion between the Sun and the
telescope. The value of the Doppler shift was retrieved from NASA's
Horizon web
interface\footnote{http://ssd.jpl.nasa.gov/horizons.cgi}. After this
procedure, the solar lines in all individual spectra were located at
the same position. The packages were then averaged to produce the
final solar atlas. We note that the whole process does not involve
measuring or correcting for the position of any of the solar lines, we
entirely rely on the telluric lines and the computation of the
relative Sun-Observer velocity. Therefore, the positions of the solar
lines is free of any assumptions about the solar atmospheric lines.

As a final data reduction step, we normalized the spectra to unity in
an iterative process. We fit a spline to the continuum that we
identified by smoothing the pixels that are not affected by solar or
telluric lines.

We provide an electronic version our IAG atlas in the online version
of this paper. Furthermore, a pdf-version of the atlas together with
the electronic files is available
\footnote{http://www.astro.physik.uni-goettingen.de/research/flux\_atlas}.

\section{Comparison of Standard Solar Flux Atlases}
\label{sect:Comparison}

\begin{figure*}
  \resizebox{0.5\hsize}{!}{\includegraphics{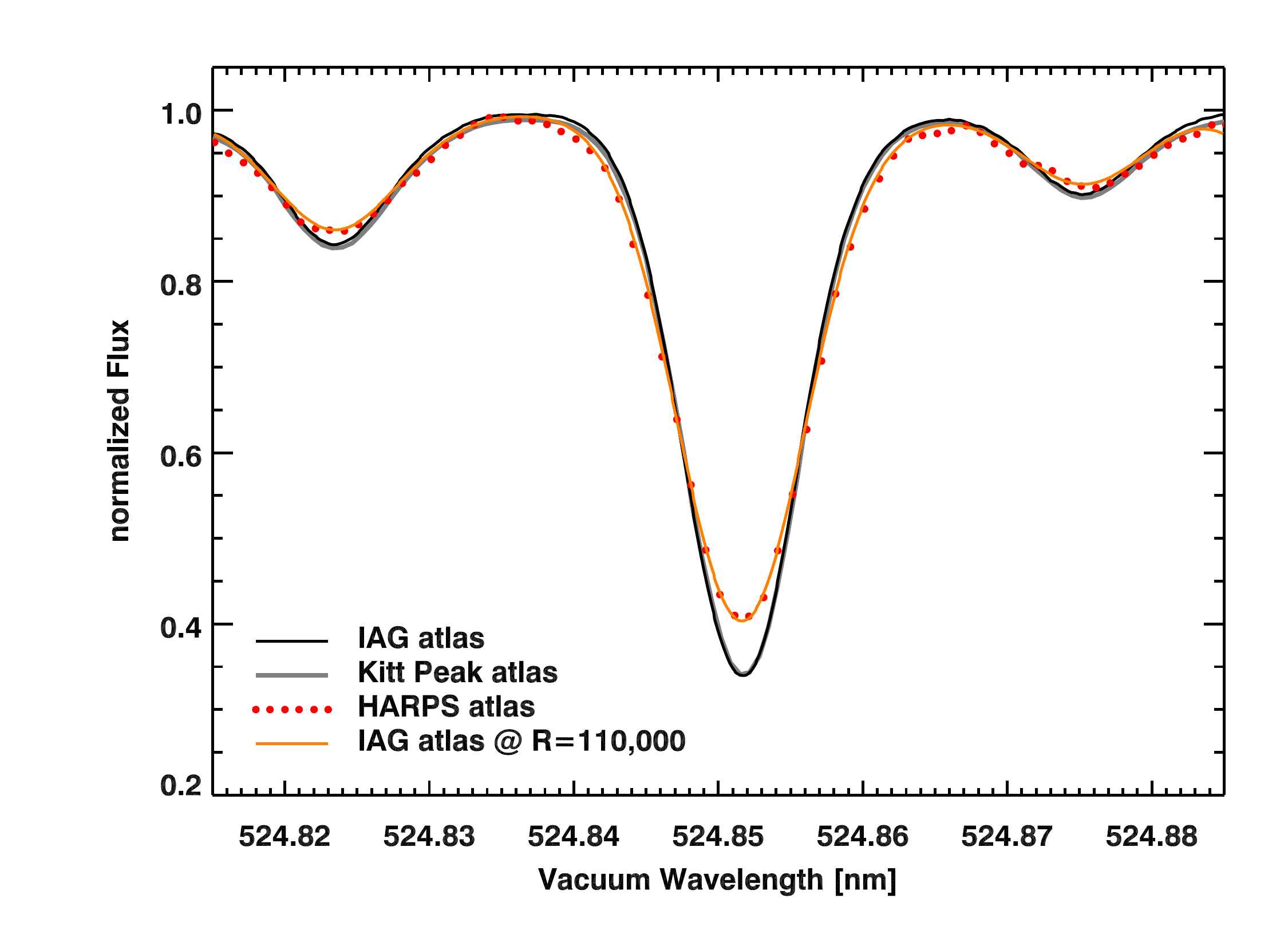}}
  \resizebox{0.5\hsize}{!}{\includegraphics{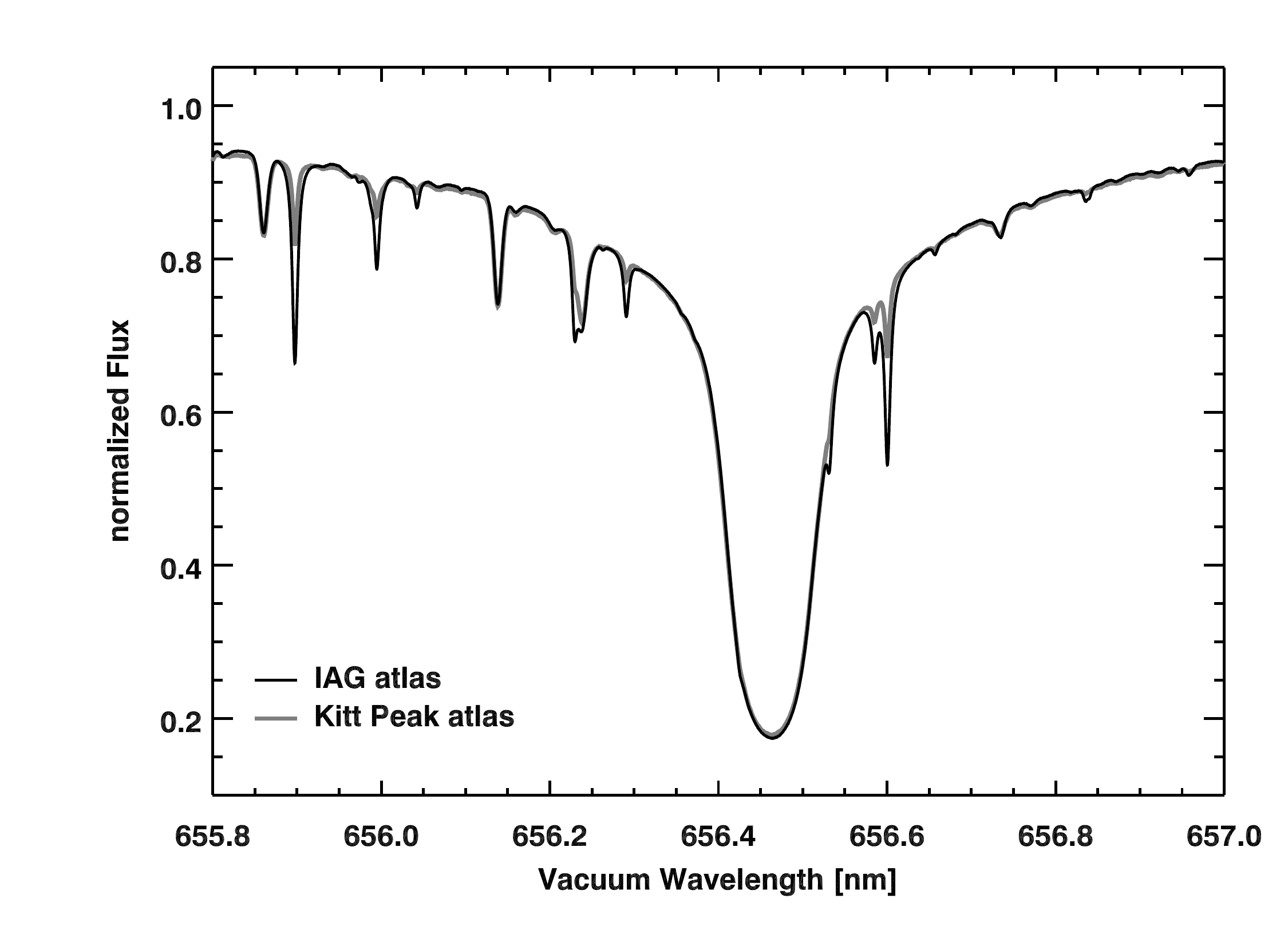}}
  \caption{Comparison of the IAG spectral atlas with the Kitt Peak FTS-atlas from
    \citet{1984sfat.book.....K} and the HARPS atlas from
    \citet{2013A&A...560A..61M}. \emph{Left:} Three absorption lines around
    524.85\,nm. To allow comparison to the HARPS atlas, the IAG atlas was
    artificially broadened using a Gaussian profile and assuming an instrument
    resolution of $R = 110,000$. \emph{Right:} Comparison in the wavelength
    range of H$\alpha$. The sharp lines are telluric absorption and mismatch
    because of different observing conditions. The HARPS atlas does not cover
    this spectral region.}
  \label{fig:Spectra}
\end{figure*}
  
\subsection{Overview}

We consider three solar flux atlases for our comparison: two atlases
taken with the FTS at Kitt Peak, and the HARPS LFC atlas. Before we
discuss differences between the atlases, we summarize how data were
taken and wavelength scales were created.

\subsubsection{Kitt Peak FTS Atlas \#1 -- \citet{1984sfat.book.....K}}

Observations for this first Kitt Peak atlas were obtained with the
McMath Solar Telescope at Kitt Peak National Observatory. In order to
increase the signal-to-noise ratio (SNR) for each scan, data were
taken in six different settings. Between 16 and 36 scans were
averaged, two different beam-splitters and two different detectors were
used \citep[see Table~1 in][]{1984sfat.book.....K}. Observations were
carried out on six different days between Nov 23, 1980 and Jun 22,
1981. The wavenumber scale correction was determined from the telluric
O$_2$ line at 688.38335\,nm in settings covering that line (\#11 and
\#13). For the redmost setting (\#15), the wavenumber scale was
corrected by aligning terrestrial lines in common with the adjacent
setting (\#13). All other settings were corrected by aligning
\emph{relatively clean solar lines in overlapping scans}
\citep{1984sfat.book.....K}. A summary of the settings is given in
Table\,\ref{tab:KittPeakScans}. The authors estimate errors of the
wavelength scale to be as large as 100\,m\,s$^{-1}$ in scan \#1,
errors in the other scans are estimated to be smaller. The scan
lengths result in a broadening of 0.2\,km\,s$^{-1}$, the value of the
middle time of observation was used for Doppler correction.

\begin{table}
  \caption{Settings and Doppler corrections used for the two Kitt Peak atlases.}
  \label{tab:KittPeakScans}
  \centering              
  \begin{tabular}{c c r r}
    \hline\hline          
    \noalign{\smallskip}
    \# Scan & Date & $\lambda$-range [nm] & $(1 + \kappa)$ \\  
     or Region\\
    \hline
    \noalign{\smallskip}
    \multicolumn{4}{c}{\citet{1984sfat.book.....K}}\\
    \noalign{\smallskip}
     1 & 1981 06 22 & 296.0 -- \phantom{1}329.9  & 1.00000250\\
     3 & 1981 06 21 & 329.9 -- \phantom{1}378.3  & 1.00000250\\
     5 & 1981 06 22 & 378.3 -- \phantom{1}402.0  & 1.00000250\\
     7 & 1980 11 23 & 402.0 -- \phantom{1}473.8  & 0.99999917\\
     9 & 1981 03 24 & 473.8 -- \phantom{1}576.5  & 0.99999950\\
    11 & 1981 03 25 & 576.5 -- \phantom{1}753.9  & 0.99999900\\
    13 & 1981 03 25 & 753.9 -- \phantom{1}999.7  & 0.99999900\\
    15 & 1981 05 11 & 999.7 -- 1300.0 & 1.00000000\\
    \noalign{\smallskip}
    \hline
    \noalign{\smallskip}
    \multicolumn{4}{c}{\citet{2011ApJS..195....6W}}\\
    \noalign{\smallskip}
     6 & 1981 06 22 & 295.8 -- \phantom{1}325.7 & 0.99999873\\
     5 & 1989 10 01 & 325.7 -- \phantom{1}401.5 & 1.00000093\\
     4 & 1989 10 01 & 401.5 -- \phantom{1}444.3 & 1.00000100\\
     3 & 1989 10 01 & 444.3 -- \phantom{1}571.3 & 1.00000046\\
     2 & 1989 10 13 & 571.3 -- \phantom{1}740.5 & 1.00000119\\
     1 & 1989 10 13 & 740.5 -- \phantom{1}925.7 & 1.00000185\\
    \noalign{\smallskip}
    \hline\hline
    \noalign{\smallskip}
    
  \end{tabular}
\end{table}

\subsubsection{Kitt Peak FTS Atlas \#2 -- \citet{2011ApJS..195....6W}}

The second Kitt Peak FTS Atlas was pieced together from scans taken
with the same instrument as used for the first Kitt Peak FTS atlas. In
addition, a version with telluric contamination removed by comparison
with a spectrum taken at solar disk center is provided. We are here
only referring to the uncorrected atlas. The specifications of the
individual scans taken for this atlas are mostly similar to the one
from \citet{1984sfat.book.....K} but a different set of scans was used
for this atlas. A major difference is the way the Doppler correction
was determined. For this second atlas, the Doppler correction was
measured empirically; line positions of solar Fe~\textsc{i} lines were
measured and corrected to the frequencies reported in the catalogue by
\citet{1994ApJS...94..221N}. This implies that both gravitational
redshift and the average velocity of the solar convective blueshift
are removed from the atlas. Details of the individual scans and their
Doppler corrections are summarized in Table\,\ref{tab:KittPeakScans}.
Electronic versions of the two Kitt Peak FTS atlases can be downloaded
at \texttt{ftp://vso.nso.edu/pub/}.

\subsubsection{HARPS LFC Atlas -- \citet{2013A&A...560A..61M}}

A solar flux atlas from the HARPS spectrograph was obtained on Nov 25,
2010 at La Silla observatory, Chile, from observations of the
Moon. The spectra were taken at a resolution of $R \sim 110,000$,
which means that in contrast to the FTS atlases, the solar lines are
not fully resolved. The wavelength solution was created using external
information from a spectrum taken with a Laser Frequency Comb
\citep[LFC;][]{2010SPIE.7735E..0TW}. The accuracy of the LFC spectral
lines is several factors higher than the one of the intrinsic FTS
solution and does not require any other correction. The spectra were
used to discover variations in pixel size and distortions of the ThAr
wavelength solution of the HARPS spectrograph. In Nov 2010, the LFC
provided light useful for wavelength calibration in the range
476--530\,nm and 534--585\,nm; the two parts fall onto different
detectors. The solar atlas is provided for this wavelengths range. The
wavelength solution is corrected for barycentric motion.

\subsection{Flux comparison}

In Fig.\,\ref{fig:Spectra}, we show in the left panel a comparison
between our new IAG atlas with the Kitt Peak FTS atlas \#1 from
\citet{1984sfat.book.....K} and with the HARPS LFC atlas from
\citet{2013A&A...560A..61M} in the region around an Fe\,\textsc{i}
line. The right panel shows the comparison around H$\alpha$ (the HARPS
LFC atlas does not cover this region). The IAG atlas and the Kitt Peak
atlas are in remarkable agreement in both regions. The sharp lines
around H$\alpha$ are telluric lines that reflect the very different
observing conditions at the two sites. In the left panel, the lower
spectral resolution of the HARPS atlas is evident. For comparison, we
include an artificially broadened version of the IAG atlas assuming an
additional Gaussian broadening according to an instrumental resolution
of $R = 110,000$. The broadened spectrum is in very good agreement
with the HARPS spectrum.

We conclude from this comparison that at the level of visual
inspection, the three atlases compare to very high degree. This shows
that the three atlases do not differ significantly although they were
taken with very different instruments. For example, one might have
expected that scattered light could influence the HARPS spectrum more
than the FTS spectra. Also, the level of the normalized continua are
very similar although rather different (standard) procedures were
applied to the unnormalized spectra. It is also apparent from
Fig.\,\ref{fig:Spectra} that the wavelength scales of the three
spectra are very similar, which shows that barycentric corrections and
conversion from air to vacuum wavelengths are consistently applied.

\subsection{Wavelength Scale Comparison}

\begin{figure*}
  \centering
  \resizebox{.9\hsize}{!}{\includegraphics{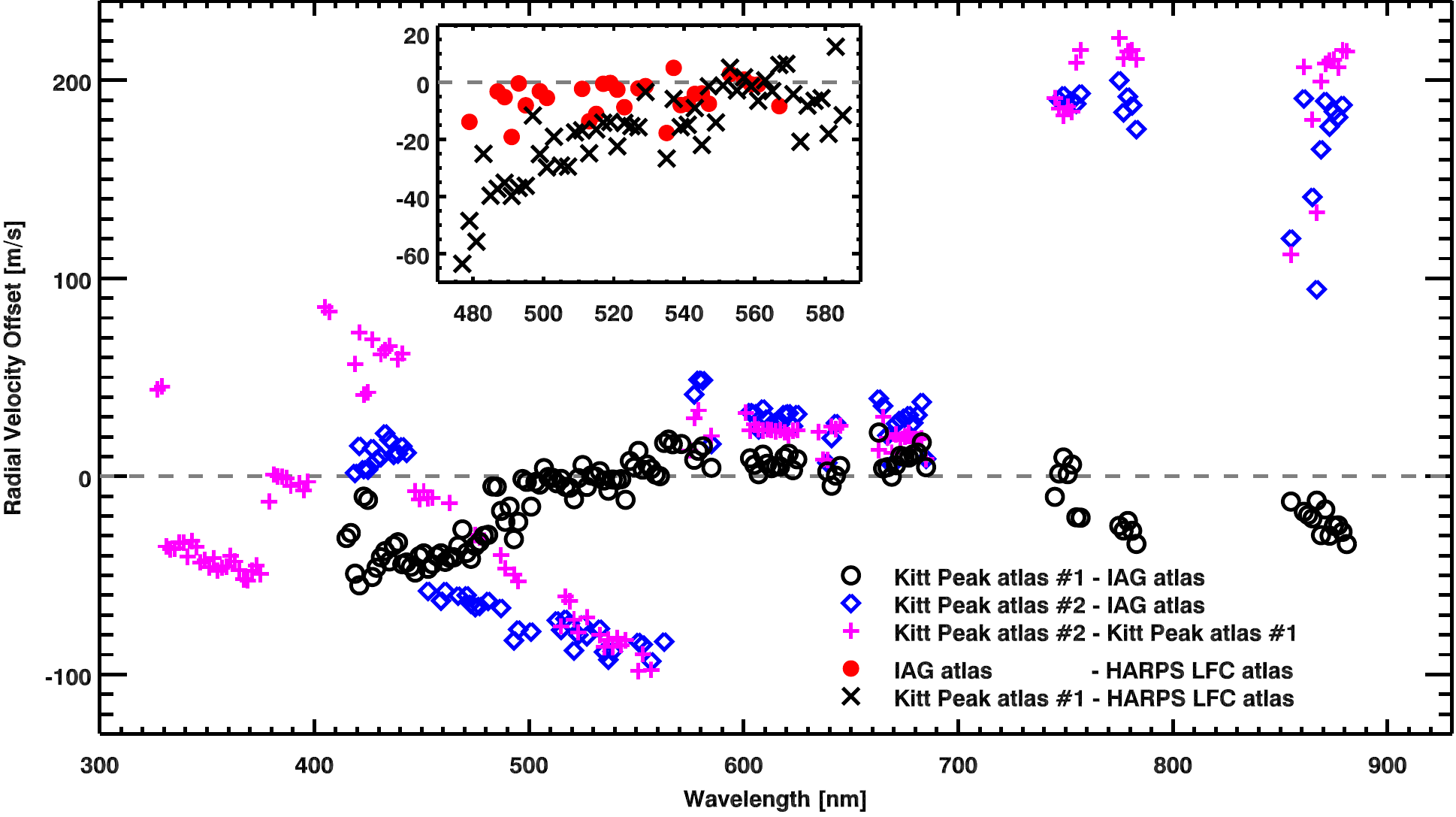}}
  \caption{Comparison of wavelength scales between different solar
    atlases --- IAG atlas: this work; Kitt Peak atlas \#1:
    \citet{1984sfat.book.....K}; Kitt Peak atlas \#2:
    \citet{2011ApJS..195....6W}; HARPS LFC atlas:
    \citet{2013A&A...560A..61M}. Each data point shows the offset
    between two atlases calculated from a cross-correlation in a 2\,nm
    wide range (see text). Comparisons between the IAG atlas and the
    two Kitt Peak atlases are shown in the main plot. Comparison
    between the HARPS atlas on the one hand and the IAG atlas and the
    Kitt Peak atlas from \citet{1984sfat.book.....K} are shown in the
    enlarged inset. Only regions with little telluric contamination
    are shown.}
  \label{fig:RadCorr}
\end{figure*}

Precision spectroscopy is investigating Doppler shifts at the level of
m\,s$^{-1}$, and future instrumentation is planned to be as precise as
10\,cm\,s$^{-1}$ \citep[e.g.,][]{2010SPIE.7735E..0FP,
  2014SPIE.9147E..26S}. As long as spectra are compared to others
taken with the same instrument and consistent calibration methods,
this precision can be achieved by securing stable calibration
procedures. As soon as spectra are compared between different
instruments, however, or if calibration procedures change, calibration
must be tied to \emph{absolute} standards for which the level of
1\,m\,s$^{-1}$ ($3 \cdot 10^{-9}$) is extremely difficult to
reach. The faintness of astronomical targets is only one part of the
problem; even with very bright sources in the laboratory, reaching an
accuracy of $\sim10^{-9}$ is a very challenging task \citep[see,
e.g.,][]{Learner:88}.

The way spectrographs are frequency- (or wavelength-) calibrated
differs significantly between a grating spectrograph and an FTS. The
three atlases used here are presumably calibrated to very high
accuracy, but there are a number of effects that can deteriorate the
final wavelength scale. In order to assess the wavelength calibrations
of the four solar atlases, we compared the two Kitt-Peak FTS atlases
and the HARPS LFC atlas against the IAG atlas, and checked consistency
by comparing the two Kitt Peak atlases with each other as well as the
HARPS LFC atlas with Kitt Peak atlas \#1. We calculated the radial
velocity offset between two atlases by cross-correlating them against
each other in chunks of 2\,nm.  The cross-correlation function (CCF)
was calculated in a range $\pm 2$\,km\,s$^{-1}$ around zero
offset. The center of the CCF was determined by fitting a second order
polynomial to the three pixels around the maximum of the CCF, i.e., at
the maximum pixel value and its two closest neighbors, and by
determining the maximum of this polynomial. All atlases were corrected
for barycentric correction, which means that spectral lines from the
solar photosphere match between all atlases but the positions of
telluric lines differ. The cross-calibration provides a measure of fit
quality between two spectra (similar to the $\chi^2$ value). We used
this information to discard regions of strong telluric contamination,
i.e., we reject regions in which the maximum value of the CCF
indicated a mismatch between two spectra.

\begin{table}
  \caption{Differential comparison of wavelength scales between solar flux atlases}
  \label{tab:comparison}
  \centering              
  \begin{tabular}{c c r r}
    \hline\hline          
    \noalign{\smallskip}
    Comparison between & $\lambda$-range  & median $\Delta \varv$ & $\sigma$\phantom{aa}  \\
                       &   [nm] & [m\,s$^{-1}$]  & [m\,s$^{-1}$] \\
    \noalign{\smallskip}
    \hline
    \noalign{\smallskip}
     Kitt Peak       \#1 & 401 -- 473 &  -38.9 &  10.0 \\
    and IAG          & 472 -- 576 &   -4.5 &  13.1 \\
                     & 576 -- 753 &    6.6 &   6.0 \\
                     & 753 -- 999 &  -23.7 &   6.3 \\
                     & 999 -- 1300 & 310.7 &   9.6 \\
    \noalign{\smallskip}
    \hline
    \noalign{\smallskip}
 Kitt Peak       \#2 & 401 -- 444 &   11.3 &   5.9 \\
    and IAG          & 444 -- 571 &  -75.0 &  10.8 \\
                     & 571 -- 740 &   28.4 &  10.1 \\
                     & 740 -- 926 &  189.5 &  47.4 \\
    \noalign{\smallskip}
    \hline
    \noalign{\smallskip}
   HARPS LFC         & 476 -- 530 &   -5.9 &   5.6 \\
     and IAG         & 534 -- 585 &   -4.1 &   6.2 \\
    \noalign{\smallskip}
    \hline \hline
    \noalign{\smallskip}
 Kitt Peak \#2       & 296 -- 326 & -605.7 &   2.6 \\
        and          & 329 -- 378 &  -42.3 &   6.0 \\
 Kitt Peak \#1       & 378 -- 401 &   -3.4 &   4.3 \\
                     & 401 -- 444 &   63.6 &  13.4 \\
                     & 444 -- 473 &  -10.3 &   2.4 \\
                     & 473 -- 571 &  -68.4 &  21.5 \\
                     & 571 -- 576 &    0.0 &   0.0 \\
                     & 576 -- 740 &   21.4 &   6.0 \\
                     & 753 -- 925 &  199.6 &  29.6 \\
    \noalign{\smallskip}
    \hline
    \noalign{\smallskip}
 HARPS LFC           & 476 -- 530 &  -27.3 &  14.3 \\
      and            & 534 -- 576 &   -6.5 &   9.4 \\
 Kitt Peak \#1       & 576 -- 585 &   -5.8 &  11.4 \\
    \noalign{\smallskip}
    \hline\hline        

  \end{tabular}
\end{table}

The results of our comparison are shown in Fig.\,\ref{fig:RadCorr} and
summarized in Table\,\ref{tab:comparison}. In Fig.\,\ref{fig:RadCorr},
we can see rather substantial differences between individual data
sets, and we can clearly distinguish the regions at which different
Doppler corrections were applied to the FTS data. We discuss the
radial velocity offsets of the two Kitt Peak atlases, and the HARPS LFC
atlas to the IAG atlas in the following.

\subsubsection{Kitt Peak atlas \#1}

The absolute difference between the IAG atlas and the Kitt Peak FTS
atlas \#1 from \citet{1984sfat.book.....K} is shown as open circles in
Fig.\ref{fig:RadCorr}. In the wavelength range 500--700\,nm, the two
wavelength scales are consistent within $\pm$10\,m\,s$^{-1}$.  A
discontinuous step of approx.\ 10\,m\,s$^{-1}$ occurs at
576\,nm. Blueward of 473\,nm, a steep gradient can be seen that
reaches down to an offset of $\sim -50$\,m\,s$^{-1}$ at 430\,nm. On
the red side of the spectrum beyond 700\,nm, a less steep gradient
appears, the offset at 850\,nm is approx.\,$-20$\,m\,s$^{-1}$. This
analysis confirms the discussion of wavelength accuracy in
\citet{1984sfat.book.....K}; best agreement is reached around the
range where the Kitt Peak atlas was anchored at the O$_2$ line at
688\,nm, and the difference between the wavelength scales increases
with distance to the anchor point. Our new atlas does not extend to
wavelengths shorter than 400\,nm but an extrapolation of the blueward
trend between 400 and 500\,nm is consistent with a maximum offset on
the order of $-100$\,m\,s$^{-1}$ around 300\,nm, as estimated in
\citet{1984sfat.book.....K}. Not shown in Fig,\,\ref{fig:RadCorr} is
the large offset at wavelengths beyond 1000\,nm, the difference there
is approx. $+300$\,m\,s$^{-1}$ (see Table\,\ref{tab:comparison}).

\subsubsection{Kitt Peak atlas \#2}

The comparison between the IAG atlas and the Kitt Peak FTS atlas \#2
of \citet{2011ApJS..195....6W} reveals interesting systematic effects
(purple rhombs in Fig.\ref{fig:RadCorr}).  In the range 570--700\,nm,
the offset between the two atlases is approximately constant at a
value of $\approx +25$\,m\,s$^{-1}$. Much larger deviations and
systematic trends occur at other regions, for example a steep gradient
between 440\,nm and 570\,nm, a large discontinuity (100\,m\,s$^{-1}$)
at 570\,nm, and offsets on the order 200\,m\,s$^{-1}$ redward of
700\,nm. We also provide offsets calculated between the two Kitt Peak
atlases (purple plus symbols in Fig.\,\ref{fig:RadCorr}). The
differences between them are as large as $600$\,m\,s$^{-1}$ at 300\,nm
(not shown in Fig.\,\ref{fig:RadCorr}) and 200\,m\,s$^{-1}$ between
700 and 900\,nm. The comparison clearly shows the consequences of
using the solar lines as wavelength standards as done in the atlas
from \citet{2011ApJS..195....6W}. The values used for correcting a
given wavelength range are essentially averages over all lines in each
range, i.e., the correction depends on the distribution of the
spectral lines, their formation depths, and the convective motion
occurring at these depths. Since the distribution of line parameters
over wavelengths is not smooth, several large jumps occur between the
individual scan regions.

\subsubsection{HARPS LFC atlas}

From a comparison between the different FTS atlases, we can learn how
their wavelength scales differ as a function of wavelength. It is not
possible, however, to determine an accurate zero-point of the
comparison at a level better than $\pm 10$\,m\,s$^{-1}$ because the
FTS atlases used above (the IAG atlas and the Kitt Peak atlas \#1) are
anchored at telluric lines. A data set that is supposed to overcome
this limitation is the LFC-calibrated solar atlas from HARPS. The
results of the comparison between the HARPS LFC atlas and the IAG
atlas, and between the HARPS LFC atlas and the Kitt Peak atlas \#1 are
shown in the inset in Fig.\,\ref{fig:RadCorr} with filled red circles
and black crosses, respectively. The HARPS LFC atlas and the IAG atlas
show a systematic offset of ($-5 \pm 5$)\,m\,s$^{-1}$ with no
significant trend (see Table\,\ref{tab:comparison}, the uncertainty is
the standard deviation of the offsets among the sample of 2\,nm wide
chunks). This is well within the estimated limits of accuracy
determined by the uncertainty of telluric line positions used for the
FTS wavelength correction. The lack of a significant trend bolsters
our assumption that the wavelength scale of the IAG VIS-atlas is
intrinsically consistent across its wavelength range and accurate
within the uncertainty of telluric line determination.

The comparison between the Kitt Peak atlas \#1 and the HARPS LFC atlas
reveals the same trend as the comparison between the IAG atlas and the
Kitt peak atlas \#1: between 534\,nm and 585\,nm, the two atlases are
consistent within uncertainties, but the average offset between
476\,nm and 530\,nm grows larger with a mean value of
$-30$\,m\,s$^{-1}$ and a steep gradient reaching to $-60$\,m\,s$^{-1}$
as shown in Fig.\,\ref{fig:RadCorr} .

\section{Convective blueshift}
\label{sect:Blueshift}

\begin{figure*}
  \resizebox{0.49\hsize}{!}{\includegraphics{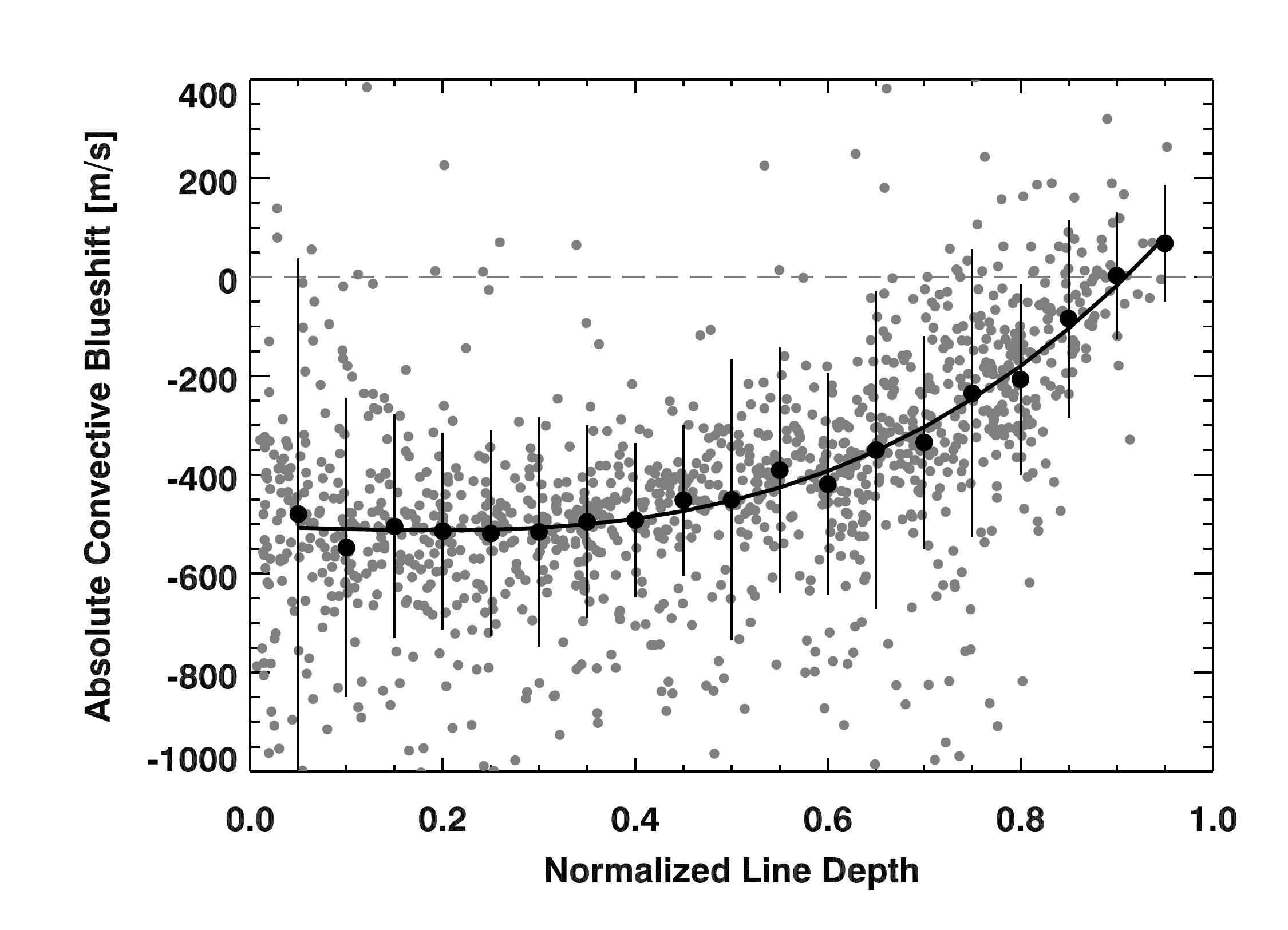}}
  \resizebox{0.49\hsize}{!}{\includegraphics{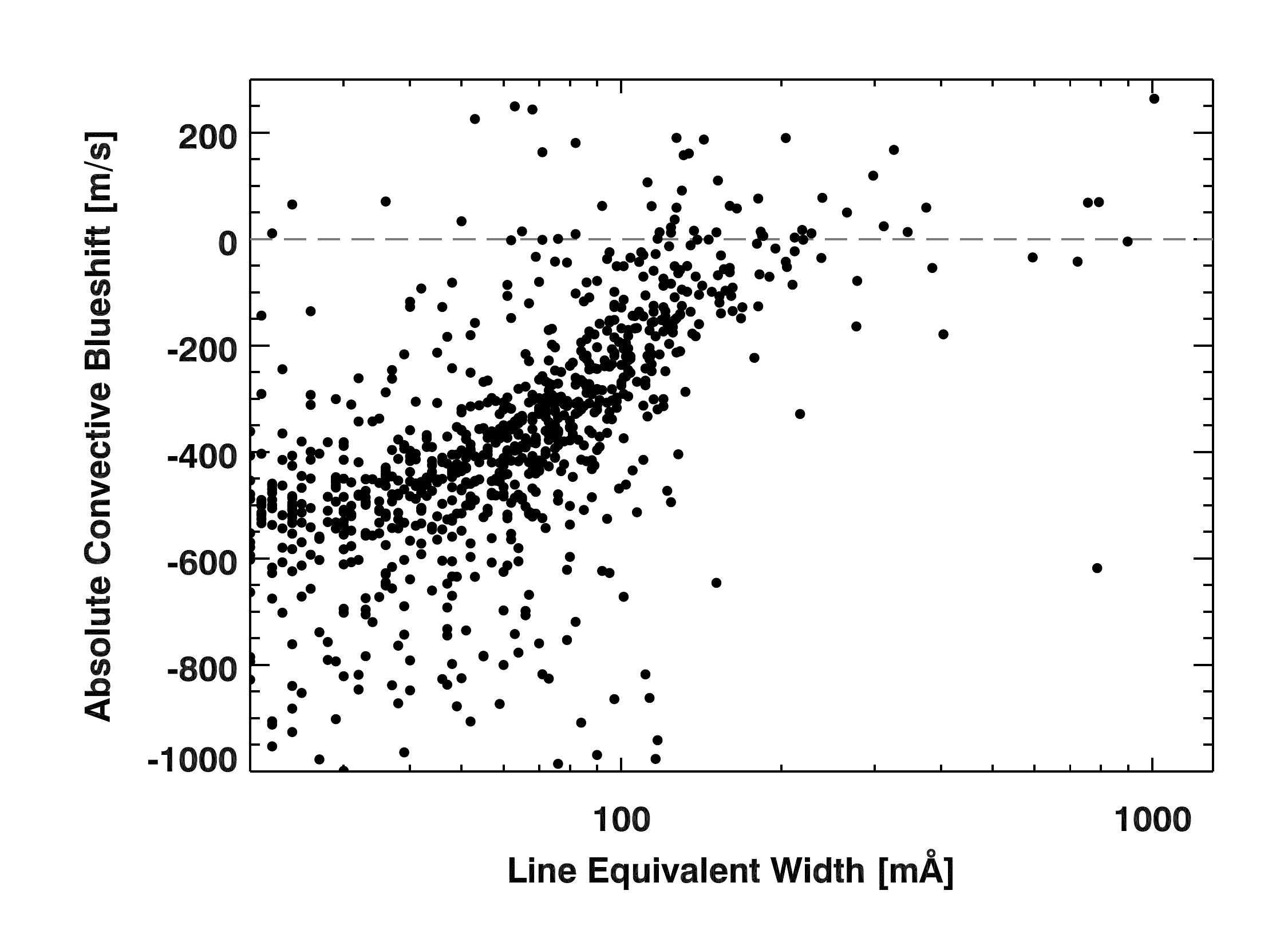}}
  \caption{Absolute convective blueshift of Fe\,\textsc{i} lines
    measured from the IAG atlas. \emph{Left:} Blueshift measurements
    for individual lines (grey points) as a function of line
    depth. The measurements are corrected for gravitational redshift
    of 636\,m\,s$^{-1}$. The median of the blueshift measurements is
    shown in bins of size 0.05, error bars show the standard deviation
    for each bin. A third order polynomial (Eq.\,\ref{eq:bspoly}) is
    fit to the binned data. \emph{Right:} Blueshift measurements as a
    function of line equivalent width as measured by
    \cite{1998A&AS..129...41A}.}
  \label{fig:blueshift}
\end{figure*}

\begin{table}
  \caption{Absolute convective blueshift as function of line depth.}
  \label{tab:blueshift}
  \centering              
  \begin{tabular}{c r c}
    \hline\hline          
    \noalign{\smallskip}
    rel.\ line depth  & mean $\Delta \varv$ & $\sigma$ \\
    & [m\,s$^{-1}$]  & [m\,s$^{-1}$] \\
    \noalign{\smallskip}
    \hline
    \noalign{\smallskip}
     0.10 & -547 & 303 \\
 0.20 & -513 & 199 \\
 0.30 & -515 & 232 \\
 0.40 & -491 & 155 \\
 0.50 & -450 & 284 \\
 0.60 & -419 & 224 \\
 0.70 & -334 & 215 \\
 0.80 & -207 & 193 \\
 0.90 &    2 & 128 \\

    \noalign{\smallskip}
    \hline\hline
  \end{tabular}
\end{table}

We make use of the IAG atlas to re-investigate the absolute convective
blueshift of solar absorption lines and its dependence on line
strength. We follow the strategy of \citet{1998A&AS..129...41A}, who
provided a comprehensive analysis of the blueshift of Fe~\textsc{i}
absorption lines as measured in the Kitt Peak atlas \#1; the authors
use lines between 400\,nm and 750\,nm. As shown above, the wavelength
solution of the Kitt Peak atlas \#1 shows some significant distortions
at the wavelength range relevant for the blueshift study of
\citet{1998A&AS..129...41A}. We therefore re-determined the blueshift
from our atlas.

For our analysis, we used the line list from
\citet{1998A&AS..129...41A} and measured positions of the 1252 lines
that are within the covered wavelength range (lines in the range
400--405\,nm are not in our atlas). The center of each line was
determined by fitting a second-order polynomial to the central
$\pm4$\,km\,s$^{-1}$ of each line, this corresponds to a range between
50\,m\AA\ and 90\,m\AA\ (Allende Prieto et al.\ used a range of
50\,m\AA\ for all lines and fitted a fourth-order polynomial; we
checked consistency of our method to the results from Allende Prieto
et al.\ by calculating the line centers in the Kitt Peak atlas \#1).
We calculated the difference between the measured line center and the
laboratory line position as provided in the catalogue of
\citet{1994ApJS...94..221N}. In Fig.\,\ref{fig:blueshift}, we show the
absolute convective blueshift of the 1252 Fe\,\textsc{i} lines as a
function of line depth (left panel) and line equivalent width (right
panel). We adopted the line equivalent widths from the compilation in
\citet{1998A&AS..129...41A}. All values of convective blueshift are
corrected for the effect of gravitational redshift, i.e., an amount of
636\,m\,s$^{-1}$ was subtracted from our measured values to provide
the absolute effect caused by the convective line shift. The
individual values are provided in electronic form as Table~A1.

The individual line shifts in Fig.\,\ref{fig:blueshift} show some
scatter caused by measurement uncertainties and blends. We binned the
individual data points in chunks of 0.05 units of normalized depth and
show the median value together with the standard deviation in the left
panel of Fig.\,\ref{fig:blueshift}; the values are also provided in
Table\,\ref{tab:blueshift} in steps of 0.1 depth units. A third order
polynomial fit to the binned data points is overplotted, the
polynomial is
\begin{equation}
  \label{eq:bspoly}
  \Delta v = -504.891 -43.7963 - 145.560 x^2 + 884.308 x^3 \ \mathrm{[m\,s}^{-1}\mathrm{]}
\end{equation}
with $x$ the relative line depth between 0.05 and 0.95.

The right panel of Fig.\,\ref{fig:blueshift} show the blueshift as a
function of line equivalent width as in Fig.\,2b of
\citet{1998A&AS..129...41A}. There are no significant differences
between the overall appearances of the two plots. We disagree,
however, with the conclusion from Allende Prieto et al.\ that lines
located in the ``plateau'' with equivalent widths larger than $\sim
100$\,m\AA\ are all blueshifted \citep[see
also][]{2009ApJ...697.1032G}. It appears that the strongest lines
occur at the shortest wavelengths where the Kitt Peak solar atlas
tends to overestimate the blueshift of the lines by several
10\,m\,s$^{-1}$. Our new determinations therefore show a
systematically lower amount of blueshift in the strongest
lines. Regardless of this systematic shift, we argue that the effect
of a ``plateau'' is not real but rather the consequence of
saturation. All spectral lines with equivalent widths larger than
$\sim 100$\,m\AA\ have line depths close to or above 90\,\%. Stronger
lines do not systematically probe higher regions in the stellar
atmosphere but gain their larger equivalent width from growing line
wings hence higher formation heights of the wings and not the center.

\section{Summary}

We introduced the IAG solar flux atlas from 405\,nm up to
2300\,nm. The atlas was taken with an FTS in only two settings, one
below approx.\ 1000\,nm and a second one above. The fidelity of our
atlas is similar to the Kitt Peak FTS atlases and we showed that the
line depths are consistent between our atlas and the Kitt Peak atlases
as well as to the atlas taken with HARPS. We summarized the
instrumental characteristics used for the different atlases and argued
that using only one setting of our entire VIS setting (405 --
1065\,nm) delivers a wavelength scale that is superior to the one in
the Kitt Peak atlases where scans covering different wavelength ranges
were stitched together. Our comparison revealed large systematic
offsets between the two Kitt Peak atlases and also between the IAG
atlas and the Kitt Peak atlases. This comes as no surprise because very
different strategies for wavelength correction were used in the Kitt
Peak atlases. For our NIR setting (1000--2300\,nm), we did not perform
a similar analysis in this paper.

Comparison of our wavelength scale to the one of the solar flux atlas
taken with the HARPS spectrograph and calibrated using an LFC shows
that the two scales are consistent within an uncertainty of only a few
m\,s$^{-1}$ as expected from the uncertainty in telluric line
positions used for calibration in the IAG atlas. We conclude that the
wavelength scale of the IAG atlas is precise and accurate to a level
of $\pm10$\,m\,s$^{-1}$. The atlas from \citet{1984sfat.book.....K} is
accurate at a 10\,m\,s$^{-1}$ level in the limited wavelength range
530--570\,nm but the wavelength accuracy deteriorates to uncertainties
as large as 60\,m\,s$^{-1}$ at 400\,nm and probably more at shorter
wavelengths as written in the original publication. The wavelength
scale of the atlas published in \citet{2011ApJS..195....6W} is
corrected for differences between solar lines and laboratory line
measurements and therefore carries only limited information about
absolute line positions. Systematic differences between the latter
atlas and the others are exceeding values of 100\,m\,s$^{-1}$ at
certain wavelengths.

We re-investigated convective blueshift in solar absorption lines as a
function of line depth and equivalent width. We found that the
systematic blueshift between 400\,nm and 500\,nm in the atlas from
\citet{1984sfat.book.....K} can lead to an underestimate of the
blueshift in the strongest solar Fe\,\textsc{i} lines that can be
found in that wavelength range. We found a steady decrease of the
blueshift from approx.\ $-550$\,m\,s$^{-1}$ to around zero shift from
very shallow (10\,\% absorption) to very deep (90\,\%) lines,
respectively. We argued that there is no plateau at very large
equivalent widths because lines with equivalent widths exceeding
$\sim100$\,m\AA\ do not actually probe higher regions of the solar
atmosphere.

Our new IAG atlas is a first attempt to provide high-accuracy and
high-fidelity benchmark spectra useful for sub-m\,s$^{-1}$ work, e.g.,
understanding the influence of convection on solar and stellar RV
measurements. For the future, we aim to provide similar data for
spatially resolved regions on the solar disk, and we are including an
LFC in our facilities to provide an accurate wavelength calibration
that overcomes the uncertainties of telluric line positions.

\begin{acknowledgements}
  We thank the referee, X.~Dumusque, for a very helpful report,
  H.-G.~Ludwig for helpful discussions, and the team at IAG for their
  support with the observations. Part of this work was supported by
  the DFG funded SFB\,963 \emph{Astrophysical Flow Instabilities and
    Turbulence} and by the ERC Starting Grant \emph{Wavelength
    Standards}, Grant Agreement Number 279347. AR acknowledges
  research funding from DFG grant RE 1664/9-1. The FTS was funded by
  the DFG and the State of Lower Saxony through the
  Gro{\ss}ger\"ateprogramm \emph{Fourier Transform Spectrograph}.
\end{acknowledgements}

%
%

\bibliographystyle{aa}
\bibliography{refs}

\end{document}